\documentclass[preprint,12pt]{elsarticle}

\usepackage{lmodern}
\usepackage{amsmath,amssymb,bm}
\usepackage{graphicx}
\usepackage{slashed}
\usepackage{booktabs}
\usepackage{xcolor}
\usepackage[colorlinks=true,linkcolor=blue,citecolor=blue,urlcolor=purple]{hyperref}
\newcommand{\arxiv}[2][]{\href{https://arxiv.org/abs/#2}{arXiv:#2\if\relax\detokenize{#1}\relax\else\ \mbox{[#1]}\fi}}
\newcommand{\doilink}[2]{\href{https://doi.org/#1}{#2}}
\newcommand{\Lp}{\Lambda\to p\,\ell^-\bar\nu_\ell}
\newcommand{\Lpmu}{\Lambda\to p\,\mu^-\bar\nu_\mu}
\newcommand{\Lpe}{\Lambda\to p\,e^-\bar\nu_e}
\newcommand{\vecp}[1]{\vec p_{#1}}
\newcommand{\GeVsq}{~\mathrm{GeV}^2}

\journal{Physics Letters B}

\begin{document}

\begin{frontmatter}

\title{Polarization in $\Lpmu$ as a probe of CP-violating new physics}

\author[lums,ncp]{Qazi Maaz Us Salam\corref{cor1}}
\ead{qazimaaz92@gmail.com}
\cortext[cor1]{Corresponding author}
\author[ncp]{Ishtiaq Ahmed}
\ead{ishtiaq.ahmed@ncp.edu.pk}

\affiliation[lums]{organization={Department of Physics, Lahore University of Management Sciences (LUMS)},
            addressline={Opposite Sector U, D.H.A},
            city={Lahore},
            postcode={54792},
            country={Pakistan}}
\affiliation[ncp]{organization={National Center for Physics},
            addressline={Shahdra Valley Road},
            city={Islamabad},
            postcode={44000},
            country={Pakistan}}

\begin{abstract}
We investigate the normal $\mu$ polarization in the hyperon decay $\Lpmu$ as a probe of new physics that violates CP, since this polarization vanishes at tree level in the Standard Model. Using an effective Lagrangian and lattice QCD form factors, we compute its sensitivity to the imaginary parts of the couplings. With the present bounds from kaon decays and the Large Hadron Collider, one coupling at a time, this sensitivity gives an effect below the Standard Model contribution from the electromagnetic final-state interaction. Only a pseudoscalar coupling with a particular phase can exceed this contribution, by several times. The final-state interaction gives the same polarization in $\Lambda$ and $\bar\Lambda$ decays, so the difference between the two decays isolates the new physics effect.
\end{abstract}

\end{frontmatter}

\section{Introduction}
\label{sec:intro}

Semileptonic decays of hyperons are an important source of information on both the weak and the strong interactions~\cite{Gaillard:1984ny,Garcia:1985xz,Cabibbo:2003cu}. In the Standard Model (SM), the strangeness-changing decays are governed by the quark-level flavor-changing charged-current (FCCC) transition $s\to u\,\ell^-\bar\nu_\ell$ and give access to the Cabibbo-Kobayashi-Maskawa (CKM) matrix element $|V_{us}|$ from the baryon sector~\cite{Cabibbo:2003cu,Bacchio:2025auj}. Beyond the SM, their rates probe non-standard scalar and tensor interactions, in a way that complements the leptonic and semileptonic decays of kaons~\cite{Chang:2014iba,Gonzalez-Alonso:2016etj,Gonzalez-Alonso:2016sip}.

Recent measurements have renewed the interest in $\Lp$. At BESIII, the $\Lambda\bar\Lambda$ pairs from $J/\psi$ decays are produced with polarized and entangled spins~\cite{Faldt:2017kgy,Perotti:2018wxm,BESIII:2018cnd}, which makes precise tests of the CP symmetry with hyperons possible~\cite{BESIII:2018cnd,BESIII:2021ypr,Du:2024jfc,Wang:2024qff}. The same collaboration measured the absolute branching ratio of the muon channel~\cite{BESIII:2021ynj} and used this polarization and entanglement to determine the axial-vector, weak-magnetism and weak-electricity couplings of the electron channel~\cite{BESIII:2026ehc}. LHCb has reported the branching ratio of the muon channel together with the ratio $R_{\mu e}=\Gamma(\Lpmu)/\Gamma(\Lpe)$, a test of lepton flavor universality (LFU)~\cite{LHCb:2025wld}, and the data samples at the proposed Super Tau-Charm Facility (STCF) should be larger by about two orders of magnitude~\cite{Achasov:2023gey,Zhou:2025skh}. Numerous theoretical studies of $\Lp$ have been carried out using a variety of approaches, including quark models~\cite{Schlumpf:1994fb,Faessler:2008ix,Ni:2026arb}, light-front dynamics~\cite{Lih:2026dcj}, QCD sum rules~\cite{Zhang:2024ick,Ahmadi:2025oal}, and lattice QCD~\cite{Bacchio:2025auj,Alexandrou:2026noh}; see Ref.~\cite{Adlarson:2026lsp} for a recent review.

Physics beyond the SM, or new physics (NP), enters $s\to u\,\ell^-\bar\nu_\ell$ at low energies through an effective Lagrangian with vector, scalar, pseudoscalar and tensor couplings $\epsilon_i$~\cite{Cirigliano:2009wk,Chang:2014iba,Gonzalez-Alonso:2016etj}, in analogy with nuclear and neutron $\beta$ decays~\cite{Bhattacharya:2011qm,Cirigliano:2013xha,Gonzalez-Alonso:2018omy,Falkowski:2020pma}. These couplings are constrained by kaon decays and by the high-energy tails of lepton production at the Large Hadron Collider (LHC)~\cite{Gonzalez-Alonso:2016etj,Allwicher:2022gkm,Allwicher:2026sdy}. Their imaginary parts violate the CP symmetry. Bounds that are linear in such an imaginary part come from the T-odd muon polarization measured by the KEK-E246 experiment in $K^+\to\pi^0\mu^+\nu_\mu$~\cite{KEK-E246:2004xuu,Abe:2006de,LopezCastro:2009da} and in $K^+\to\mu^+\nu_\mu\gamma$~\cite{KEKE246:2003yie}. The hadronic matrix elements are parametrized by form factors, which are the source of the theoretical uncertainty. For $\Lambda\to p$, the vector and axial-vector form factors have recently been computed in lattice QCD at physical quark masses~\cite{Bacchio:2025auj}, followed by the scalar and tensor form factors, which were used to constrain real scalar and tensor couplings with the ratio $R_{\mu e}$~\cite{Alexandrou:2026noh}. The pseudoscalar form factor follows from the axial Ward identity, so all hadronic inputs needed for an NP analysis of $\Lp$ are now available.

In our previous work~\cite{Salam:2026vns}, we calculated the longitudinal and transverse components of the charged-lepton polarization in $\Lp$ with lattice form factors, as well as the normal component $P_N$, which is generated in the SM by the electromagnetic final-state interaction (FSI). This component is proportional to the triple product $\vec s_\ell\cdot(\vecp{\ell}\times\vecp{\Lambda})$, a quantity that changes sign under naive time reversal. At tree level, a nonzero $P_N$ needs a relative phase between amplitudes, which makes it a direct probe of CP-violating couplings, in the same way as the transverse muon polarization in $K_{\mu3}$ decays~\cite{KEK-E246:2004xuu,Abe:2006de,LopezCastro:2009da}, the transverse electron polarization in neutron decay~\cite{Kozela:2009ni,Kozela:2011mc}, the $\tau$ normal polarization in $b$-hadron decays~\cite{Ivanov:2017mrj,Ajaltouni:2019log}, and T-odd observables in $b\to s\,\ell^+\ell^-$ decays~\cite{Aliev:2002ux,Aliev:2002nv}. T violation is not necessary for such a correlation, however, because the FSI can also produce it~\cite{Khriplovich:1967sop,Okun:1967ww,Efrosinin:2000yv,Braguta:2002jy}. For $\Lpmu$, the FSI gives $\langle P_N^\mu\rangle_{\rm SM}=-(1.06\pm0.15)\times10^{-3}$~\cite{Salam:2026vns}, the SM value against which any CP-violating signal in this observable must be compared.

In this study, we compute $P_N$ in $\Lp$ with complex scalar, pseudoscalar, tensor and right-handed couplings, taking the vector, axial-vector, scalar and tensor form factors from lattice QCD and the pseudoscalar one from the axial Ward identity. We obtain its sensitivity to the imaginary parts of the couplings and combine it with the present bounds from KEK-E246, from the ratio $R_K=\Gamma(K\to e\nu)/\Gamma(K\to\mu\nu)$ and from the LHC. Unlike the rates, which are largely insensitive to the pseudoscalar coupling~\cite{Chang:2014iba}, the normal polarization depends on it directly. We show that the decays $K\to\mu\nu$ leave one direction of the complex pseudoscalar coupling unconstrained, along which the normal muon polarization in $\Lpmu$ can exceed the SM contribution by several times. We also show that the NP contribution can be separated from the FSI through its $q^2$ dependence and through the comparison of $\Lambda$ and $\bar\Lambda$ decays. We compute the squared amplitudes using the Mathematica package \textsc{FeynCalc}~\cite{Mertig:1990an,Shtabovenko:2016sxi,Shtabovenko:2020gxv,Shtabovenko:2023idz}, as in our previous works~\cite{Salam:2026qzj,Salam:2026vns}, with a general spin vector of the charged lepton.

The rest of this paper is organized in the following way. In Sec.~\ref{sec:lagrangian}, we introduce the effective Lagrangian and the hadronic matrix elements. In Sec.~\ref{sec:pn}, we study the normal polarization and its sensitivity to NP. In Sec.~\ref{sec:bounds}, we examine the present bounds and the largest effects they allow, and we conclude in Sec.~\ref{sec:concl}. The expressions for the sensitivities are given in \ref{app:closed}.

\section{Effective Lagrangian and hadronic matrix elements}
\label{sec:lagrangian}

In this section, we present the effective Lagrangian and the hadronic matrix elements. The most general low-energy $s\to u\,\ell^-\bar\nu_\ell$ interaction with left-handed neutrinos is~\cite{Cirigliano:2009wk,Chang:2014iba,Gonzalez-Alonso:2016etj,Alexandrou:2026noh}
\begin{align}
\mathcal{L}_{\rm eff}=-\frac{G_FV_{us}}{\sqrt2}\Big[&\bar\ell\gamma_\mu(1-\gamma_5)\nu_\ell\;\bar u\gamma^\mu\big(1-(1-2\epsilon_R)\gamma_5\big)s\nonumber\\
&+\epsilon_S\,\bar\ell(1-\gamma_5)\nu_\ell\;\bar u s-\epsilon_P\,\bar\ell(1-\gamma_5)\nu_\ell\;\bar u\gamma_5 s\nonumber\\
&-\epsilon_T\,\bar\ell\sigma_{\mu\nu}(1-\gamma_5)\nu_\ell\;\bar u\sigma^{\mu\nu}(1-\gamma_5)s\Big]+\text{h.c.},
\label{eq:Lnp}
\end{align}
where $G_F$ is the Fermi constant and $\sigma^{\mu\nu}=\frac i2[\gamma^\mu,\gamma^\nu]$. The sign of $\epsilon_T$ follows Refs.~\cite{Chang:2014iba,Alexandrou:2026noh}. The couplings $\epsilon_i$ are complex and are taken at $\mu=2$~GeV in the $\overline{\rm MS}$ scheme, and the overall factor $1+\epsilon_L+\epsilon_R$ of the vector and axial-vector currents is absorbed in $V_{us}$~\cite{Chang:2014iba,Gonzalez-Alonso:2016etj}. The couplings depend on the lepton flavor, and $\epsilon_i$ refers to the lepton under consideration.

For the vector and axial-vector currents, we use the standard decomposition~\cite{Weinberg:1958ut,Bacchio:2025auj}
\begin{align}
\langle p(p')|\bar u\,\gamma^\mu(1-\gamma_5)\,s|\Lambda(p)\rangle=\bar u_p(p')\Big[&\gamma^\mu\big(f_1-g_1\gamma_5\big)-\frac{i\sigma^{\mu\nu}q_\nu}{m_\Lambda}\big(f_2-g_2\gamma_5\big)\nonumber\\
&+\frac{q^\mu}{m_\Lambda}\big(f_3-g_3\gamma_5\big)\Big]u_\Lambda(p),
\label{eq:ffVA}
\end{align}
with $q=p-p'$, and for the scalar and tensor currents~\cite{Alexandrou:2026noh} and the pseudoscalar current
\begin{align}
\langle p|\bar us|\Lambda\rangle&=f_S\,\bar u_p u_\Lambda,\qquad
\langle p|\bar u\gamma_5s|\Lambda\rangle=g_P\,\bar u_p\gamma_5u_\Lambda,\nonumber\\
\langle p|\bar u\,i\sigma^{\mu\nu}s|\Lambda\rangle&=\bar u_p\Big[i\sigma^{\mu\nu}T_1+\frac{\gamma^{[\mu}p^{\nu]}}{m_\Lambda}T_2+\frac{\gamma^{[\mu}p'^{\nu]}}{m_p}T_3+\frac{p^{[\mu}p'^{\nu]}}{m_\Lambda m_p}T_4\Big]u_\Lambda ,
\label{eq:ffST}
\end{align}
where $a^{[\mu}b^{\nu]}=a^\mu b^\nu-a^\nu b^\mu$ and all form factors depend on $q^2$. For $f_i$ and $g_i$, we use the lattice results of Ref.~\cite{Bacchio:2025auj} with the uncertainties of our previous work~\cite{Salam:2026vns}, which add to the lattice errors and correlations a $2\%$ uncertainty on the charges, the shift obtained by setting $g_1/f_1$ to its experimental value, and the shift obtained by scaling all radii by $1.5$. For $f_S$ and $T_{1,\dots,4}$, we use the lattice results of Ref.~\cite{Alexandrou:2026noh}, listed in Table~\ref{tab:inputs}. No lattice result is available for $g_P$, which we obtain from the axial Ward identity,
\begin{equation}
g_P(q^2)=\frac{(m_\Lambda+m_p)\,g_1(q^2)-q^2g_3(q^2)/m_\Lambda}{m_s+m_u}.
\label{eq:gP}
\end{equation}

\begin{table}[t]
\centering
\caption{Scalar and tensor form factors~\cite{Alexandrou:2026noh} and quark masses~\cite{ParticleDataGroup:2026mpi}.}
\label{tab:inputs}
\begin{tabular}{lcccccc}
\toprule
 & $f_S$ & $T_1$ & $T_2$ & $T_3$ & $T_4$ & \\
\midrule
$F(0)$ & $2.296(48)$ & $0.7000(46)$ & $-1.71(10)$ & $1.361(67)$ & $-0.804(75)$ & \\
$\langle r^2\rangle m_\Lambda^2$ & $12.14(94)$ & $4.42(43)$ & $22.0(3.2)$ & $17.0(2.9)$ & $23.6(4.8)$ & \\
\midrule
 & \multicolumn{3}{c}{$m_s=92.9\pm0.7$~MeV} & \multicolumn{3}{c}{$m_u=2.16\pm0.07$~MeV}\\
\bottomrule
\end{tabular}
\end{table}

The form factors in Table~\ref{tab:inputs} are used in the form $F(q^2)=F(0)\big(1+\langle r^2\rangle q^2/6\big)$ given in Ref.~\cite{Alexandrou:2026noh}, and the quark masses are the $\overline{\rm MS}$ values at $2$~GeV. With these masses, Eq.~(\ref{eq:gP}) gives $g_P(0)=17.7(8)$, which rises to $19.8(9)$ at $q^2_{\max}=(m_\Lambda-m_p)^2$.

To compute the polarization, we apply the spin projector $\frac12(1+\gamma_5\slashed{S})$ to the spinor of the charged lepton~\cite{Salam:2026vns}. We keep the spin four-vector $S$ general, with $S^2=-1$ and $S\cdot p_\ell=0$, through the traces, and insert the longitudinal, transverse and normal spin vectors afterwards. The normal spin vector requires the Levi-Civita tensor, whose sign convention we fix with a trace evaluated with explicit Dirac matrices. In the SM limit, the unpolarized rate agrees with Eq.~(32) of Ref.~\cite{Bacchio:2025auj}, and the longitudinal and transverse polarizations agree with our previous work~\cite{Salam:2026vns}. For real couplings, Ref.~\cite{Alexandrou:2026noh} writes $R_{\mu e}=R_{\mu e}^{\rm SM}(1+r_S\epsilon_S+r_T\epsilon_T)$. With the lattice hadron masses used there, we find $R_{\mu e}^{\rm SM}=0.1736$ and $r_S=1.270$, in agreement with their $0.1735(98)$ and $1.271(29)$. For the tensor coupling, we find $r_T=2.862$.
\section{Normal polarization and its sensitivity to new physics}
\label{sec:pn}

In this section, we discuss the normal polarization and its sensitivity to NP. The normal polarization is defined as
\begin{equation}
P_N(q^2)=\frac{d\Gamma(+)/dq^2-d\Gamma(-)/dq^2}{d\Gamma(+)/dq^2+d\Gamma(-)/dq^2},
\label{eq:PN}
\end{equation}
where $\pm$ refers to the spin of the charged lepton along $\hat e_N=\vecp{\ell}\times\vecp{\Lambda}/|\vecp{\ell}\times\vecp{\Lambda}|$, with the momenta in the dilepton rest frame, as in Ref.~\cite{Salam:2026vns}. Its average $\langle P_N\rangle$ over the full phase space is the ratio of the separately integrated numerator and denominator of Eq.~(\ref{eq:PN}).

The normal polarization measures the triple product $\vec s_\ell\cdot(\vecp{\ell}\times\vecp{\Lambda})$. Since the lattice form factors are real, at tree level it is proportional to the imaginary parts of products of the couplings. It therefore vanishes in the SM and for real $\epsilon_i$, and for small couplings it is linear in ${\rm Im}\,\epsilon_S$, ${\rm Im}\,\epsilon_P$ and ${\rm Im}\,\epsilon_T$. We have verified these properties numerically. It is worth mentioning here that a complex right-handed coupling does not generate $P_N$. The coupling $\epsilon_R$ only changes the relative phase of the vector and axial-vector hadronic amplitudes, and the resulting T-odd term has opposite signs for the two proton helicities, so it cancels in the sum over the proton spin. Furthermore, scalar, pseudoscalar and tensor interactions produce a charged lepton of the opposite helicity to that of the SM current, so their interference with the SM in the rate is proportional to $m_\ell$. A spin component perpendicular to the lepton momentum, however, measures the interference of the two helicity amplitudes directly, and $P_N$ is therefore not suppressed by $m_\ell$.

After averaging over $q^2$, we find
\begin{align}
\langle P_N^\mu\rangle&=-0.694(39)\,{\rm Im}\,\epsilon_S+0.332(13)\,{\rm Im}\,\epsilon_P+0.400(13)\,{\rm Im}\,\epsilon_T,\nonumber\\
\langle P_N^e\rangle&=-1.245(70)\,{\rm Im}\,\epsilon_S+0.586(23)\,{\rm Im}\,\epsilon_P+0.772(21)\,{\rm Im}\,\epsilon_T,
\label{eq:PNsens}
\end{align}
where the couplings refer to the muon and to the electron, respectively. The uncertainties combine in quadrature the uncertainties of the vector and axial-vector form factors described above, those of $f_S$ and $T_{1,\dots,4}$ in Table~\ref{tab:inputs} with the correlations given in Ref.~\cite{Alexandrou:2026noh}, and that of $m_s$ for $\epsilon_P$. The first source dominates for all three couplings. In agreement with the helicity argument above, the sensitivities of the electron are not suppressed, and they are even larger than those of the muon. Expressions for the sensitivities at fixed $q^2$ are given in \ref{app:closed}.

\begin{figure}[t]
\centering
\includegraphics[width=0.62\textwidth]{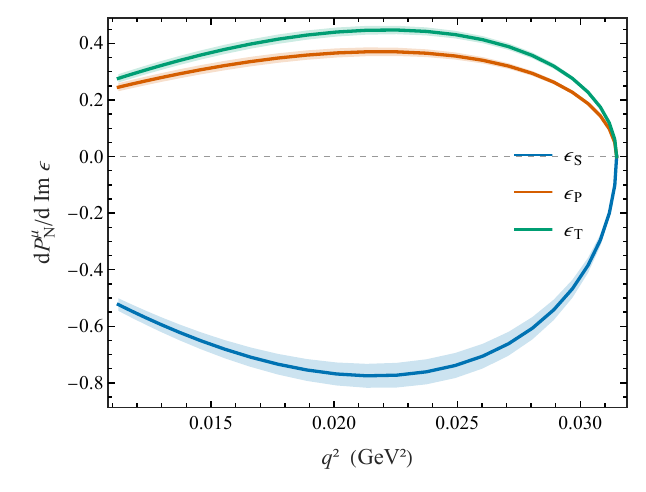}
\caption{Sensitivity of $P_N^\mu$ to the imaginary parts of the couplings.}
\label{fig:sens}
\end{figure}

In Fig.~\ref{fig:sens}, we present the $q^2$ dependence of the sensitivities of the muon, with the uncertainties of the form factors represented by shaded bands. It can be seen that all three sensitivities have the same shape. They grow from threshold, reach their largest values near $q^2\simeq0.022\GeVsq$, and vanish at $q^2_{\max}=(m_\Lambda-m_p)^2$. The ratios between them change by less than $13\%$ over the full $q^2$ range, so the $q^2$ dependence alone can hardly distinguish the three couplings.

For a complex pseudoscalar coupling alone, the amplitude is linear in $\epsilon_P$, so the squared amplitude contains only the terms $1$, $\epsilon_P$, $\epsilon_P^*$ and $|\epsilon_P|^2$, and the integrated rate and the normal polarization take the exact form
\begin{equation}
\frac{\Gamma}{\Gamma_{\rm SM}}=1-0.172\,{\rm Re}\,\epsilon_P+0.178\,|\epsilon_P|^2,\qquad
\langle P_N^\mu\rangle=\frac{0.332\,{\rm Im}\,\epsilon_P}{\Gamma/\Gamma_{\rm SM}},
\label{eq:closedP}
\end{equation}
which we use in Sec.~\ref{sec:bounds}.

\section{Present bounds and maximal effects}
\label{sec:bounds}

In this section, we discuss the present bounds on the imaginary parts of the muonic couplings and the largest normal polarization they allow, taking one coupling at a time.

\emph{Scalar coupling.} The KEK-E246 experiment measured the T-odd polarization of the muon perpendicular to the decay plane in $K^+\to\pi^0\mu^+\nu_\mu$, ${\rm Im}\,\xi=-0.0053\pm0.0071\pm0.0036$, i.e., $|{\rm Im}\,\xi|<0.016$ at $90\%$ confidence level (CL)~\cite{KEK-E246:2004xuu,Abe:2006de}. Here $\xi=f_-/f_+$ is the ratio of the $K\to\pi$ form factors, and a scalar coupling acts as an additional contribution to $f_-$~\cite{LopezCastro:2009da}. We computed the polarization of the muon in $K_{\mu3}$ with the same Lagrangian and the matrix elements of Ref.~\cite{Gonzalez-Alonso:2016etj}, averaged over the Dalitz plot, and found ${\rm Im}\,\xi=22.7\,{\rm Im}\,\epsilon_S$, hence $|{\rm Im}\,\epsilon_S|<7.1\times10^{-4}$. This agrees with the corresponding estimate $7.5\times10^{-4}$ of Ref.~\cite{LopezCastro:2009da}, which used $m_s-m_u=100$~MeV.

\emph{LHC.} Scalar, pseudoscalar and tensor operators do not interfere with the SM in $pp\to\mu\nu$ at high transverse mass, so the LHC bounds $|\epsilon_i|$, i.e., the real and imaginary parts alike. From the single-operator bounds of Ref.~\cite{Allwicher:2022gkm}, obtained from the LHC data at $\sqrt s=13$~TeV, with tree-level matching and running to $2$~GeV~\cite{Gonzalez-Alonso:2016etj}, we obtain $|\epsilon_S|,|\epsilon_P|<1.46\times10^{-2}$ and $|\epsilon_T|<1.88\times10^{-3}$ at $95\%$ CL.

\emph{Pseudoscalar coupling.} The coupling $\epsilon_P$ does not contribute to $K_{\ell3}$, but it enters the leptonic decay $K\to\mu\nu$ with a chiral enhancement~\cite{Gonzalez-Alonso:2016etj}. With NP only in the muon channel,
\begin{equation}
R_K=\frac{\Gamma(K\to e\nu)}{\Gamma(K\to\mu\nu)}=\frac{R_K^{\rm SM}}{|1-\kappa\,\epsilon_P|^2},\qquad \kappa=\frac{m_K^2}{m_\mu(m_s+m_u)}=24.3.
\label{eq:RK}
\end{equation}
The measured value $R_K=(2.488\pm0.009)\times10^{-5}$~\cite{ParticleDataGroup:2026mpi}, dominated by NA62~\cite{NA62:2012lny}, and the SM prediction $R_K^{\rm SM}=(2.47653\pm0.00034)\times10^{-5}$~\cite{Boyle:2026lrz}, as compared in Ref.~\cite{Allwicher:2026sdy}, give $|1-\kappa\,\epsilon_P|^2=0.9954\pm0.0036$. Since ${\rm Im}\,\epsilon_P$ enters only quadratically, this constraint is a thin ring around the circle $|1-\kappa\,\epsilon_P|=1$, which passes through $\epsilon_P=0$~\cite{Gonzalez-Alonso:2016etj}. For a purely imaginary $\epsilon_P$, $|1-\kappa\,\epsilon_P|^2=1+\kappa^2({\rm Im}\,\epsilon_P)^2$ can only be larger than one, whereas the data lie $1.3\sigma$ below one. Taking $\Delta\chi^2=3.84$ with respect to the best fit, ${\rm Im}\,\epsilon_P=0$, we obtain $|{\rm Im}\,\epsilon_P|<2.5\times10^{-3}$. Along the circle, $\epsilon_P=(1-e^{-i\varphi})/\kappa$, the rate of $K\to\mu\nu$ is unchanged, and ${\rm Im}\,\epsilon_P$ is limited by the LHC bound, which allows $|\varphi|<20^\circ$, i.e., $\epsilon_P=(2.6+14.4\,i)\times10^{-3}$ at the edge. Along this direction, the rate of $\Lpmu$ changes by less than $0.05\%$, Eq.~(\ref{eq:closedP}), so this direction cannot be excluded by the decay rates of hyperons or by $R_{\mu e}$, whereas $P_N$ is linear in ${\rm Im}\,\epsilon_P$.

\begin{table}[t]
\centering
\caption{Present bounds and the largest $|\langle P_N^\mu\rangle|$ they allow.}
\label{tab:max}
\small
\setlength{\tabcolsep}{4pt}
\begin{tabular}{llccc}
\toprule
Coupling & Source (CL) & Bound & $10^3\,|\langle P_N^\mu\rangle|_{\max}$ & Ratio to SM\\
\midrule
${\rm Im}\,\epsilon_S$ & KEK-E246 ($90\%$) & $7.1\times10^{-4}$ & $0.490\pm0.028$ & $0.46$\\
${\rm Im}\,\epsilon_P$, ${\rm Re}\,\epsilon_P=0$ & $R_K$ ($95\%$) & $2.5\times10^{-3}$ & $0.847\pm0.033$ & $0.80$\\
${\rm Im}\,\epsilon_T$ & LHC ($95\%$) & $1.88\times10^{-3}$ & $0.752\pm0.025$ & $0.71$\\
${\rm Im}\,\epsilon_P$, circle & $R_K$, LHC ($95\%$) & $1.44\times10^{-2}$ & $4.78\pm0.19$ & $4.5$\\
${\rm Im}\,\epsilon_R$ & & & $0$ & $0$\\
\bottomrule
\end{tabular}
\end{table}

In Table~\ref{tab:max}, we present these bounds, where the circle refers to $\epsilon_P$ on $|1-\kappa\,\epsilon_P|=1$, together with the largest $|\langle P_N^\mu\rangle|$ they allow, computed exactly at the bound, and its ratio to the SM contribution $|\langle P_N^\mu\rangle_{\rm SM}|=1.06\times10^{-3}$. The uncertainties come from the sensitivities, Eq.~(\ref{eq:PNsens}), and the sign of $\langle P_N^\mu\rangle$ is fixed by that of the imaginary part. It can be seen that scalar and tensor couplings, and a pseudoscalar coupling with a generic phase, can produce at most $|\langle P_N^\mu\rangle|$ between $0.49\times10^{-3}$ and $0.85\times10^{-3}$, below the SM contribution. Since the SM contribution is calculable, it can be subtracted. Its uncertainty, $1.5\times10^{-4}$~\cite{Salam:2026vns}, corresponds to $|{\rm Im}\,\epsilon_S|\simeq2\times10^{-4}$, $|{\rm Im}\,\epsilon_P|\simeq5\times10^{-4}$ and $|{\rm Im}\,\epsilon_T|\simeq4\times10^{-4}$, between three and six times below the present bounds. Along the direction of the complex $\epsilon_P$ that $K\to\mu\nu$ does not constrain, $|\langle P_N^\mu\rangle|$ can reach $4.78\times10^{-3}$, which is $4.5$ times the SM contribution.

The radiative decay $K^+\to\mu^+\nu_\mu\gamma$ is sensitive to this phase, because its inner-bremsstrahlung and structure-dependent amplitudes interfere~\cite{Wu:1996hi}. We find $P_T=C\,\kappa\,{\rm Im}\,\epsilon_P$, with $C$ between $0.02$ and $0.075$ in the KEK-E246 selection, depending on the form factors~\cite{OKA:2019gav} and the cuts, assuming uniform acceptance. At the edge of the LHC bound this gives $|P_T|$ between $0.7\times10^{-2}$ and $2.6\times10^{-2}$, within the $95\%$ range of the measured $(-0.64\pm1.85\pm0.10)\times10^{-2}$~\cite{KEKE246:2003yie}, so this measurement does not yet improve on the LHC bound. This direction also requires $|{\rm Im}\,\epsilon_S|\ll|{\rm Im}\,\epsilon_P|$, since for a quark current of one chirality $\epsilon_S=\pm\epsilon_P$~\cite{Gonzalez-Alonso:2016etj} and the KEK-E246 bound on ${\rm Im}\,\epsilon_S$ applies~\cite{Bezrukov:2003jc}.

\begin{figure}[t]
\centering
\includegraphics[width=0.58\textwidth]{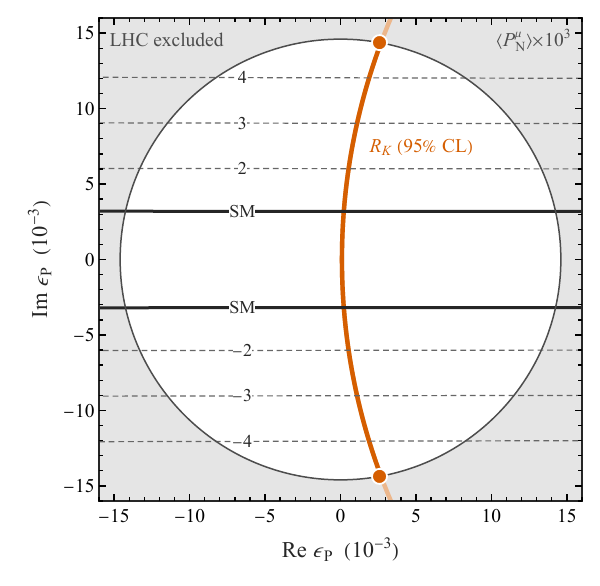}
\caption{Complex plane of $\epsilon_P$ with the present bounds and the contours of $\langle P_N^\mu\rangle$.}
\label{fig:cplx}
\end{figure}

In Fig.~\ref{fig:cplx}, we present the complex plane of $\epsilon_P$ in units of $10^{-3}$. The grey region is excluded by the LHC, the orange band is the $95\%$ CL interval of $|1-\kappa\,\epsilon_P|^2$ from $R_K$, and the dots mark the largest ${\rm Im}\,\epsilon_P$ allowed by both. On the imaginary axis, this band extends to $|{\rm Im}\,\epsilon_P|=2.0\times10^{-3}$, slightly below the bound in Table~\ref{tab:max}, which follows from the one-parameter $\Delta\chi^2$ for a purely imaginary $\epsilon_P$. The dashed lines are contours of $\langle P_N^\mu\rangle$, labeled by its value in units of $10^{-3}$, and the solid lines labeled SM mark $|\langle P_N^\mu\rangle|=1.06\times10^{-3}$, the size of the SM contribution. It can be seen that the contours are nearly horizontal, since $P_N$ measures the imaginary part and the real part enters only through the small change of the rate, Eq.~(\ref{eq:closedP}). Near the origin, the $R_K$ band lies between the two SM lines, while along the circle it crosses them and reaches the largest values of Table~\ref{tab:max}.

\begin{figure}[t]
\centering
\includegraphics[width=0.62\textwidth]{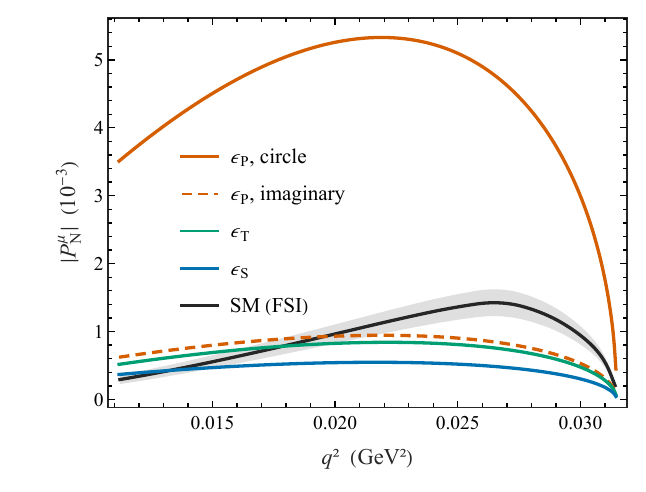}
\caption{Largest $|P_N^\mu|$ allowed by the present bounds and the SM contribution.}
\label{fig:max}
\end{figure}

In Fig.~\ref{fig:max}, we present the $q^2$ dependence of the largest $|P_N^\mu|$ allowed by the bounds of Table~\ref{tab:max}, together with the SM contribution of the FSI and its uncertainty~\cite{Salam:2026vns}. It can be seen that the two have a different $q^2$ dependence. The NP contributions are largest near $q^2\simeq0.022\GeVsq$, and their values at $q^2=0.015$ and $0.027\GeVsq$ differ by less than $6\%$. The FSI contribution, in contrast, is largest near $q^2\simeq0.027\GeVsq$, where the proton-muon relative velocity is small, and at $q^2=0.015\GeVsq$ it is $2.5$ times smaller. A measurement in bins of $q^2$ can therefore distinguish an NP contribution from the FSI, in addition to the subtraction of the calculated SM value. Moreover, the CP-conjugate decay $\bar\Lambda\to\bar p\,\mu^+\nu_\mu$ has the same FSI contribution and an NP contribution of opposite sign, so the difference of the normal polarizations in the two decays isolates the NP contribution without a subtraction of the SM value~\cite{Datta:2003mj}.

It is important to mention here that the measurement of $P_N$ requires the analysis of the muon spin. If the muon spin is determined from the direction of the electron in the muon decay, whose angular distribution, integrated over the electron energy, has the asymmetry $\alpha=\xi/3\simeq1/3$ with the Michel parameter $\xi\simeq1$~\cite{ParticleDataGroup:2026mpi}, the statistical uncertainty is $\sigma(P_N)\simeq\sqrt3/(\alpha\sqrt N)$ for $N$ analyzed decays, if the muon keeps its polarization until it decays. A precision of $10^{-3}$ therefore requires about $3\times10^7$ analyzed decays of $\Lpmu$, well beyond the samples expected at the STCF~\cite{Achasov:2023gey,Zhou:2025skh}. Negative muons, however, keep only about one sixth of their polarization after their capture into a muonic atom in the material where they stop~\cite{Mann:1961zz,Kadono:1986zz}, which increases this number by a factor of about $36$, whereas positive muons, as in the CP-conjugate decay, usually keep their full polarization~\cite{Brewer:2026yfo}. The electron channel has larger sensitivities, Eq.~(\ref{eq:PNsens}), and an FSI contribution of only order $10^{-5}$~\cite{Salam:2026vns}, but requires the measurement of the electron polarization.

\section{Conclusions}
\label{sec:concl}

In this work, we have studied the T-odd normal polarization of the muon in $\Lpmu$, which proceeds through the quark-level transition $s\to u\,\mu^-\bar\nu_\mu$, as a probe of CP-violating NP. We used the general low-energy effective Lagrangian with complex scalar, pseudoscalar, tensor and right-handed couplings, and the vector, axial-vector, scalar and tensor form factors from lattice QCD. Our analysis shows that the normal polarization is linear in the imaginary parts of the scalar, pseudoscalar and tensor couplings, with sensitivities of order one that are not suppressed by the lepton mass, and that the right-handed coupling does not contribute.

Furthermore, we have combined these sensitivities with the present bounds from KEK-E246, from the ratio of the leptonic kaon decay widths and from the LHC. Taking one coupling at a time, we find that the largest NP contribution to the normal muon polarization allowed for scalar and tensor couplings, and for a pseudoscalar coupling with a generic phase, lies between about one half and four fifths of the SM contribution of the electromagnetic FSI. In addition, we have shown that the decays $K\to\mu\nu$ leave unconstrained one direction in the complex plane of the pseudoscalar coupling, along which the NP contribution to the normal muon polarization can be $4.5$ times larger than the SM contribution, while the rate of $\Lpmu$ changes by less than $0.05\%$. This direction requires a scalar coupling much smaller than the pseudoscalar one, and it can also be tested with the T-odd polarization in $K^+\to\mu^+\nu_\mu\gamma$. Moreover, the NP contribution takes nearly equal values at $q^2=0.015$ and $0.027\GeVsq$, whereas the FSI contribution is $2.5$ times larger at the higher value, so a measurement in bins of $q^2$ can distinguish the two. Finally, the CP-conjugate decay $\bar\Lambda\to\bar p\,\mu^+\nu_\mu$ has the same FSI contribution and an NP contribution of opposite sign, so the difference of the normal polarizations in the two decays isolates the NP contribution without a subtraction of the SM value.

Overall, our results demonstrate that the normal polarization in $\Lpmu$ is complementary to kaon decays in the search for CP-violating NP. The SM contribution is calculable, and its uncertainty corresponds to couplings between three and six times below the present bounds. A measurement at the level of $10^{-3}$ requires measuring the muon spin in samples of order $10^7$ decays, or about $36$ times more for negative muons, which lose most of their polarization when they stop. Such samples may become possible at future high-intensity hyperon facilities.

\section*{Acknowledgments}

Q.~M.~U.~Salam is very thankful to the National Center for Physics (NCP), Islamabad, where he is a hosted researcher, for its hospitality during his stay while this manuscript was written.

\appendix

\section{Expressions for the sensitivities}
\label{app:closed}

In this appendix, we give expressions for the sensitivities of $P_N$ at fixed $q^2$, from which Fig.~\ref{fig:sens} and Eq.~(\ref{eq:PNsens}) follow. We use $\Sigma=m_\Lambda+m_p$, $\Delta=m_\Lambda-m_p$, $s_\pm=(m_\Lambda\pm m_p)^2-q^2$, $r=m_\ell^2/q^2$, and the helicity form factors of Ref.~\cite{Bacchio:2025auj},
\begin{align}
F_+&=f_1+\frac{q^2f_2}{m_\Lambda\Sigma},& F_\perp&=f_1+\frac{\Sigma f_2}{m_\Lambda},& F_0&=f_1+\frac{q^2f_3}{m_\Lambda\Delta},\nonumber\\
G_+&=g_1-\frac{q^2g_2}{m_\Lambda\Delta},& G_\perp&=g_1-\frac{\Delta g_2}{m_\Lambda},& G_0&=g_1-\frac{q^2g_3}{m_\Lambda\Sigma}.
\label{eq:helff}
\end{align}
In this notation, the SM rate, Eq.~(32) of Ref.~\cite{Bacchio:2025auj}, is
\begin{equation}
\frac{d\Gamma_{\rm SM}}{dq^2}=\frac{G_F^2|V_{us}|^2\sqrt{s_+s_-}}{192\pi^3m_\Lambda^3}\,(1-r)^2\,s_+B,
\label{eq:rateSM}
\end{equation}
with
\begin{align}
B={}&\frac{3r}{2}\Big[\Delta^2F_0^2+\frac{\Sigma^2s_-}{s_+}G_0^2\Big]+\frac{r+2}{2}\Big[\Delta^2G_+^2+\frac{\Sigma^2s_-}{s_+}F_+^2\Big]\nonumber\\
&+(r+2)\,q^2\Big[G_\perp^2+\frac{s_-}{s_+}F_\perp^2\Big].
\label{eq:Bsm}
\end{align}

For $S=\pm(0,\hat e_N)$, the scalar products of $S$ with all momenta vanish in the dilepton rest frame. The spin-dependent part of $|\mathcal M|^2$ then contains only the Levi-Civita tensor, through ${\rm Tr}[\gamma_5\gamma^\alpha\gamma^\beta\gamma^\gamma\gamma^\delta]=-4i\varepsilon^{\alpha\beta\gamma\delta}$ with $\gamma_5=i\gamma^0\gamma^1\gamma^2\gamma^3$ and $\varepsilon^{0123}=+1$. We write $\langle a,b,c,d\rangle=\varepsilon^{\alpha\beta\gamma\delta}a_\alpha b_\beta c_\gamma d_\delta$. The factor $i$ is the reason for only the imaginary parts of the couplings contribute. For the scalar coupling, the part of the lepton trace that contributes is $4i\langle p_\ell,S,p_\nu,\cdot\rangle$, and the hadron trace with the SM current gives a vector $h_1p+h_2p'$ with $h_1+h_2=4\Sigma F_+$, in which $f_3$ cancels. With $p_\nu=p-p'-p_\ell$, the term of $|\mathcal M|^2$ proportional to ${\rm Im}\,\epsilon_S$, summed over the baryon spins and without the factor $G_F^2|V_{us}|^2/2$, is
\begin{equation}
-32\,{\rm Im}\,\epsilon_S\,f_S\,\Sigma F_+\,\langle p_\ell,S,p,p'\rangle,\qquad \langle p_\ell,S,p,p'\rangle=\pm\sqrt{q^2}\,|\vecp{\ell}|\,|\vecp{\Lambda}|\sin\theta,
\end{equation}
where $\theta$ is the angle between $\vecp{\ell}$ and $\vecp{\Lambda}$. For the pseudoscalar coupling, the hadron trace selects the axial-vector form factors and gives $4\Delta G_+$, and the sign of $\epsilon_P$ in Eq.~(\ref{eq:Lnp}) reverses the overall sign. The tensor coupling requires the same steps with more terms. We take the difference of the two spin states, average over $\cos\theta$ with $\langle\sin\theta\rangle=\pi/4$, and divide by the sum of the two spin states in the SM, which is $\frac{16}{3}(1-r)\,s_+B$ in the same normalization. This gives
\begin{align}
\frac{dP_N}{d\,{\rm Im}\,\epsilon_S}&=-C\,f_S\,\Sigma F_+,&
\frac{dP_N}{d\,{\rm Im}\,\epsilon_P}&=C\,g_P\,\Delta G_+,\nonumber\\
\frac{dP_N}{d\,{\rm Im}\,\epsilon_T}&=C\,\mathcal T,&
C&=\frac{3\pi}{4B}\sqrt{\frac{q^2s_-}{s_+}},
\label{eq:sensclosed}
\end{align}
with
\begin{align}
\mathcal T&=(1-r)\,(F_\perp X_F+G_\perp X_G)+r\,(F_0Y_F+G_0Y_G),\nonumber\\
X_F&=-4\Delta T_1+\frac{2s_-}{m_\Lambda}T_2+\frac{2s_-}{m_p}T_3,\qquad
Y_F=-4\Delta\Big(T_1-T_2+T_3+\frac{s_+T_4}{2m_\Lambda m_p}\Big),\nonumber\\
X_G&=4\Sigma T_1-\frac{2(m_\Lambda^2-m_p^2+q^2)}{m_\Lambda}T_2-\frac{2(m_\Lambda^2-m_p^2-q^2)}{m_p}T_3,\qquad Y_G=4\Sigma T_1.
\label{eq:calT}
\end{align}
We have checked that these expressions agree identically with the result of our full calculation. Eq.~(\ref{eq:PNsens}) follows from the average with the SM rate of Eq.~(\ref{eq:rateSM}),
\begin{equation}
\frac{\langle P_N\rangle}{{\rm Im}\,\epsilon_i}=\int_{m_\ell^2}^{q^2_{\max}}dq^2\,\frac{d\Gamma_{\rm SM}}{dq^2}\,\frac{dP_N}{d\,{\rm Im}\,\epsilon_i}\Big/\int_{m_\ell^2}^{q^2_{\max}}dq^2\,\frac{d\Gamma_{\rm SM}}{dq^2}.
\end{equation}

\end{document}